\documentclass[twocolumn,superscriptaddress,amsmath,amssymb,aps, prl]{revtex4-2}

\usepackage{graphicx}
\usepackage{dcolumn}
\usepackage{bm}
\usepackage{color}

\usepackage{titlesec}
\titleformat{\section}[block]{\bfseries}{\thesection}{0.5em}{}
\titlespacing*{\section}{0pt}{1.5ex plus .2ex minus .2ex}{0.5em}

\begin{document}


\title{Generation of Ultra-Broadband Frequency Comb in Strongly Bistable Nonlinear Magnonic Resonator}

\author{Yu Jiang}
\affiliation{ 
    Department of Electrical and Computer Engineering, Northeastern University, Boston, MA 02115, USA
}

\author{Vasyl Tyberkevych}
\affiliation{ 
    Department of Physics, Oakland University, Rochester, MI 48309, USA
}

\author{Yizhong Huang}
\affiliation{ 
    Department of Electrical and Computer Engineering, Northeastern University, Boston, MA 02115, USA
}

\author{Zixin Yan}
\affiliation{ 
    Department of Electrical and Computer Engineering, Northeastern University, Boston, MA 02115, USA
}

\author{Amin Pishehvar}
\affiliation{ 
    Department of Electrical and Computer Engineering, Northeastern University, Boston, MA 02115, USA
}

\author{Andrei Slavin}
\affiliation{ 
    Department of Physics, Oakland University, Rochester, MI 48309, USA
}

\author{Xufeng Zhang}
\email{xu.zhang@northeastern.edu}
\affiliation{ 
    Department of Electrical and Computer Engineering, Northeastern University, Boston, MA 02115, USA
}
\affiliation{ 
    Department of Physics, Northeastern University, Boston, MA 02115, USA
}

\date{\today}

\begin{abstract}
Magnonic frequency combs (MFCs) offer a promising route to compact, energy-efficient platforms for on-chip coherent microwave signal generation and processing. Conventional on-chip comb generation typically relies on nonlinear resonators supporting a series of equidistant, low-loss resonances driven by a strong monochromatic signal, resulting in fixed comb spacing defined by the resonator’s free spectral range (FSR). Here we introduce and experimentally demonstrate a fundamentally different mechanism for ultrabroadband MFC generation using a highly nonlinear miniaturized magnonic resonator. The small resonator volume, combined with a slow-wave transducer, yields high intra-resonator power density, driving the system deep into the bistable regime where parametric excitation of propagating spin waves facilitates comb formation. Our approach yields more than 350 comb lines spanning a 450~MHz bandwidth, with spacing continuously tunable via a two-tone external drive, representing an order-of-magnitude enhancement over prior reports while operating at relatively low power. The platform is ultra-compact (4–6 orders of magnitude smaller in size than conventional YIG sphere resonators), fully scalable, and highly tunable, enabling precise control of comb properties through magnetic bias and pump manipulation. These results establish a new paradigm for frequency comb technology, unlocking transformative opportunities in microwave signal processing, neuromorphic computing, and precision sensing.
\end{abstract}


\maketitle



Frequency combs—spectral structures of evenly spaced lines—are foundational tools across diverse scientific and technological domains \cite{Diddams2010,Kippenberg2011,Diddams2020,Fortier2024}. 
Initially developed in photonic systems \cite{Udem2002,Hall2006,DelHaye2007,Chang2022}, 
they have enabled breakthroughs in precision spectroscopy \cite{Picque2018,Weichman2018}, 
optical clocks \cite{Udem2002,Ludlow2015}, high-speed communications \cite{Pasquazi2018,Moss2024,Yang2023} and quantum technology \cite{2016_Science_Reimer,2020_Science_Chou,2024_PRX_Dalvit}. More recently, the concept has been extended to magnonic systems, exploiting their exceptional tunability and compact form factors \cite{Pirro2021,Flebus2023,Han2024,Wang2024Networks,Zheng2023}. This emerging approach, known as the magnonic frequency comb (MFC) \cite{2024_NP_Wang_EPMFC,2023_PRL_Xu_MagnomechanicalResonator,2022_APL_Hula_SpinWaveFC,2025_APLQuantum_Kani_SqueezedComb,2023_FundResearch_Xiong,2023_PRA_Liu_TwoToneDrive,2025_PRA_Wang_MechanicalMFC,2022_PRL_Wang_Twisted,2024_NanoLett_Li_AsymmetricMFC,2021_PRL_Wang_nonlinearMagnonSkyrmion,2021_AdvEngMatt_Sun_StrainModulation,2024_PRB_Liu_SyntheticFerrimagnets}, offers strong potential for complete on-chip integration without off-chip components, while providing widely reconfigurable operation bands and controllable comb-line spacing.

Conventional on-chip frequency comb generation relies on a nonlinear resonator supporting a series of equidistant, narrow-linewidth resonances \cite{DelHaye2007,Stern2018,Yang2024,Pasquazi2018,Chang2022,Flower2024}. When driven by a high-power monochromatic signal, cascaded four-wave mixing provides gain to offset resonance losses, producing combs with fixed spacing equal to the resonator’s free spectral range (FSR). In magnonics, combs can also be generated by driving a single nonlinear magnonic resonance with multiple microwave tones \cite{2022_APL_Hula_SpinWaveFC}, producing a well-defined comb without the need for multiple resonances and thereby offering significantly enhanced in-situ tunability and reconfigurability. Although this approach has attracted considerable interest, existing demonstrations are limited in bandwidth and comb-line count \cite{2024_NP_Wang_EPMFC,2023_PRL_Xu_MagnomechanicalResonator,2022_APL_Hula_SpinWaveFC,2025_APLQuantum_Kani_SqueezedComb,2023_FundResearch_Xiong,2023_PRA_Liu_TwoToneDrive,2025_PRA_Wang_MechanicalMFC,2022_PRL_Wang_Twisted,2024_NanoLett_Li_AsymmetricMFC,2021_PRL_Wang_nonlinearMagnonSkyrmion,2021_AdvEngMatt_Sun_StrainModulation,2024_PRB_Liu_SyntheticFerrimagnets}, primarily due to challenges in device design, inefficient magnon excitation, and insufficient nonlinear enhancement.

\begin{figure*}[t]
\includegraphics[width=0.9  \linewidth]{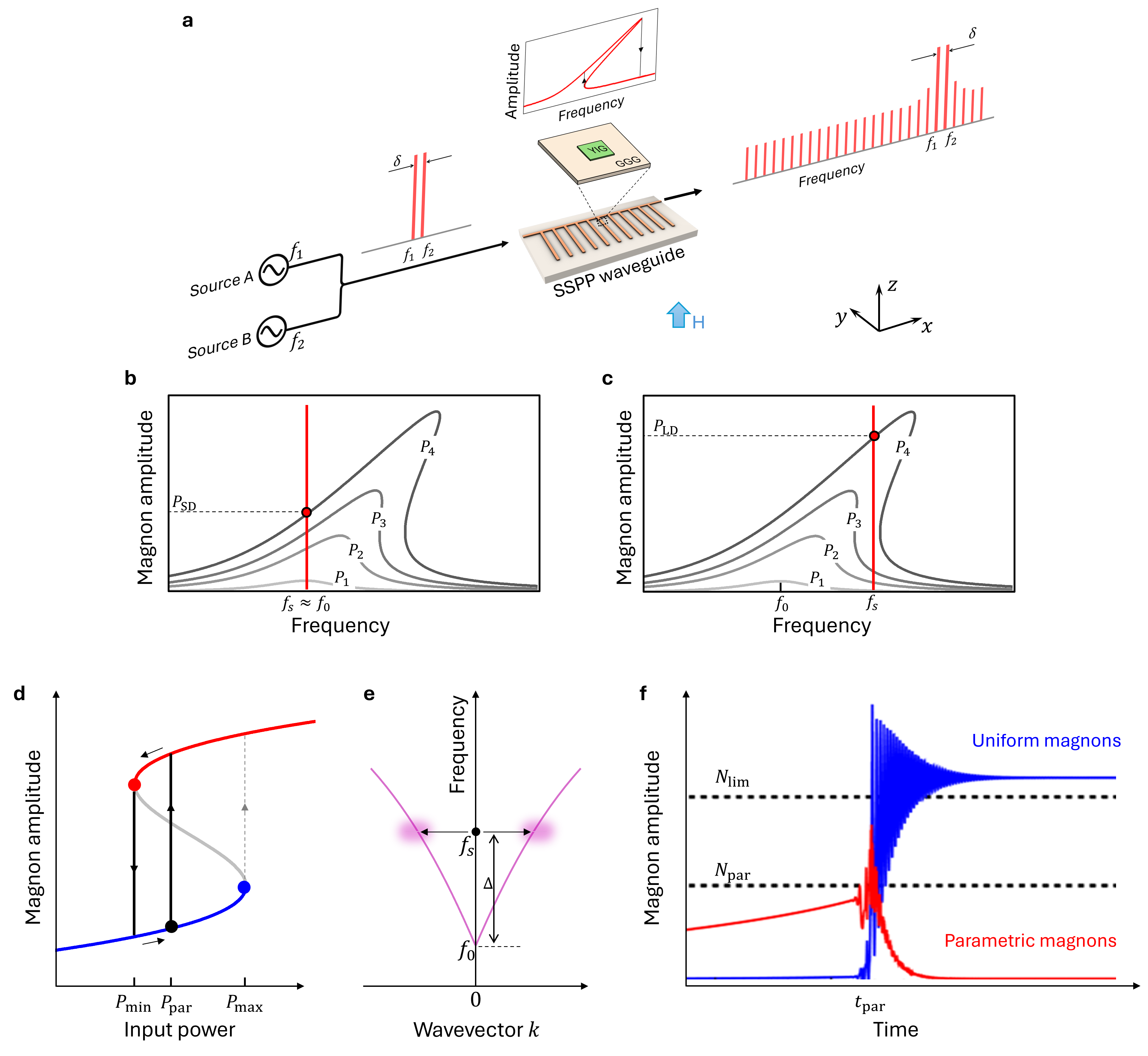}
\caption{\textbf{Concept and operating principles of the nonlinear magnonic microresonator.}
    \textbf{a}, Schematic of the device architecture: a normally magnetized YIG microresonator flip-bonded onto a slow-wave microwave waveguide supporting spoof surface plasmon polaritons. The device is driven by a two-tone microwave pump consisting of two frequencies $f_s$ and $f_s+\delta$, which exhibits an time-varying envelope at the beat frequency $\delta$. The device output is analyzed using a spectrum analyzer.
    \textbf{b,c}, Magnon resonance spectra under different input powers. Red dots mark the achievable powers for small detuning ($f_s \approx f_0$) and large detuning ($f_s \gg f_0$) for $P_4$.
    \textbf{d}, Bistability and hysteresis in the magnon power at large detuning observed when sweeping the input power. Transitions occur from the lower (upper) branch to the upper (lower) branch at $P_\mathrm{max}$ ($P_\mathrm{min}$) during upward (downward) sweeps. In the presence of parametric pumping of spin waves, $P_\mathrm{max}$ is reduced to $P_\mathrm{par}$.
    \textbf{e}, Magnon dispersion (purple curves) and parametric pumping at $f_s > f_0$. Off-resonant driving of uniform magnon mode at $f_s$ excites incoherent large-wavevector spin-wave magnon pairs (purple shaded regions) at the same frequency.
    \textbf{f}, Temporal evolution of uniform magnon mode amplitude (blue) and parametric spin-wave magnons (red). $N_\mathrm{par}$ denotes the threshold for parametric excitation; $N_\mathrm{lim}$ marks the level where the uniform magnon mode shifts to $f_s$.
    }
\label{fig1}
\end{figure*}

To overcome these constraints, we introduce a miniaturized magnonic resonator based on a patterned 200-nm-thick yttrium iron garnet (YIG) film, driven by a two-tone signal to enable ultra-broadband MFC generation. Unlike previous demonstrations, our approach leverages strong detuning of the driving tones from the linear resonance to push the resonator deep into a large-amplitude bistable regime, where four-wave parametric excitation of propagating spin waves occurs at the pump frequencies. Cross-mode nonlinear interactions between these spin waves and the uniform mode induce rapid oscillations of the uniform mode amplitude between two bistable states (Fig.~\ref{fig1}a), giving rise to MFC formation. This approach does not rely on a series of equidistant resonances, allowing flexible control of comb-line spacing, which is set by the frequency difference between the two driving components. The small resonator volume, combined with a highly efficiency slow-wave microwave transducer, ensures high power density in the resonator---accordingly significantly enhanced nonlinearity and a wide bistability loop---leading to low-threshold, high-density MFC generation. Compared with prior work, our experiments demonstrate at least an order-of-magnitude improvement in comb bandwidth and line count, along with highly tunable spacing, and achieve a threefold enhancement over a parallel study \cite{2025_PRL_Xu} that relies on a series of resonances and therefore exhibits fixed comb spacing. This on-chip MFC platform provides a scalable and tunable solution for next-generation electronics, unlocking new regimes of spin-wave dynamics for advanced functionalities in computing, sensing, and quantum technologies.

\section{Large-Amplitude Nonlinear Magnonic Response}

Our MFC device (Fig.~\ref{fig1}a) comprises a perpendicularly magnetized yttrium iron garnet (YIG) thin-film microresonator driven by a two-tone, periodically varying microwave signal to generate the comb. Compared with conventional millimeter-scale YIG spheres, our thin-film microresonator ($0.2 \times 50 \times 50~\mu\mathrm{m}^3$) reduces the device volume by 4–6 orders of magnitude, substantially increasing power density for a given microwave drive and thereby enhancing comb generation. Integration with a spoof surface plasmon polariton (SSPP) waveguide further amplifies the driving field through the slow-wave effect \cite{2024_PRL_SlowWave}, ensuring efficient power delivery to the uniform magnon mode of the YIG resonator. The intrinsic frequency $\omega_0 = 2\pi f_0$ of this mode is tuned slightly below the cutoff frequency (around 10.1 GHz) of the SSPP mode, where field enhancement is strongest.

When driven by a detuned microwave signal at $\omega_s = 2\pi f_s \gg \omega_0$, the system supports a uniform magnon mode ($k = 0$) and spin waves with nonzero wavevectors ($\mathbf{k}$), both exhibiting strong nonlinear dynamics. Their amplitudes—$a$ for the uniform magnon mode and $c_{\mathbf{k}}$ for the spin-wave mode—evolve according to:
\begin{align}
\frac{da}{dt} + \Gamma a &=
- i\omega_0^{N} a
- i\omega_{M}\!\left(\sum_{\mathbf{k}} c_{\mathbf{k}} c_{-\mathbf{k}}\right) a^{*} \nonumber \\
&\quad + iF_{s}\!\left(1 + e^{-i2\pi \delta t}\right) e^{-i\omega_{s} t},
\label{eq:4a} \\[6pt]
\frac{dc_{\mathbf{k}}}{dt} + \Gamma_{\mathbf{k}} c_{\mathbf{k}} &=
- i\omega_{\mathbf{k}}^{N} c_{\mathbf{k}}
- i\omega_{M} a^{2} c_{-\mathbf{k}}^{*}.
\label{eq:4b}
\end{align}
\noindent where the mode frequencies of the uniform magnon mode ($\omega_0^N$) and the spin-wave modes ($\omega_{\mathbf{k}}^N$), nonlinearly shifted by self- and cross-mode interactions, are given by:
\begin{align}
\omega_0^{N} &= \omega_{0} + \omega_{M}|a|^{2} 
+ 2\omega_{M}\!\sum_{\mathbf{k}} |c_{\mathbf{k}}|^{2}, 
\label{eq:nonlinearUni} \\[6pt]
\omega_{\mathbf{k}}^{N} &= \omega_{\mathbf{k}} + 2\omega_{M}|a|^{2} 
+ 2\omega_{M}\!\sum_{\mathbf{k}} |c_{\mathbf{k}}|^{2},
\label{eq:nonlinearSW}
\end{align}
where $\omega_0$ ($\omega_\mathbf{k}$) is the intrinsic angular frequency of the uniform (spin wave) magnons, and $\Gamma$ ($\Gamma_{\mathbf{k}}$) denote the damping rate of the uniform (spin wave) magnon mode. For simplicity and without loss of generality, we assume $\Gamma_{\mathbf{k}} = \Gamma$. Here, $\omega_M$ represents the nonlinear coupling coefficient, $F_s$ and $\omega_s$ are the amplitude and angular frequency of the microwave drive, and $\delta$ denotes the frequency difference between the two microwave components, which sets the comb line spacing of the resulting MFC.

Equations~\eqref{eq:4a}--\eqref{eq:4b} describe two important effects. First, there is parametric interaction between the uniform resonator mode $a$ and the pairs of plane spin wave modes $c_{\pm \mathbf{k}}$, described by the term $\omega_{M}\left(\sum_{\mathbf{k}}c_{\mathbf{k}}c_{-\mathbf{k}}\right)a^{*}$ 
in Eq.~\eqref{eq:4a} and the term  $\omega_{M}a^{2}c_{-\mathbf{k}}^{*}$ in Eq.~\eqref{eq:4b}. Second, these equations along with Eqs.~\eqref{eq:nonlinearUni} and \eqref{eq:nonlinearSW} describe mutual nonlinear frequency shifts leading to nonlinear renormalization of the resonant frequency of the uniform resonator mode and spin waves.

Under the influence of the self-nonlinearity, the uniform resonator mode exhibits a hardening Duffing-type resonance, meaning its frequency shifts upward with increasing power, and becomes bistable at sufficiently high power (e.g., $P_4$ in Figs.~\ref{fig1}b and \ref{fig1}c). Consequently, the achievable power of the uniform magnonic mode ($P_\mathrm{SD}$) is limited when the microwave drive $F_{s}$ is applied at small detuning $\Delta = f_s - f_0 \approx 0$ (Fig.~\ref{fig1}b). For MFC generation, large magnon power is required, which can only be achieved with a driving at large detuning (Fig.~\ref{fig1}c). Note that Fig.~\ref{fig1}c illustrates a modest detuning $\Delta \approx \Gamma$ for clarity. In practical devices, $\Delta$ can greatly exceed $\Gamma$, enabling much higher magnon power. Importantly, Fig.~\ref{fig1}c also shows that, due to bistability, a high input power (e.g., $P_4$) is required to reach the high-amplitude regime (the upper branch of the nonlinear magnonic resonance curve in the bistability range). This power dependence is clearly demonstrated by the hysteresis loop in Fig.~\ref{fig1}d, which shows that the system transitions to the upper branch at $P_\mathrm{max}$ during an up-sweep. However, $P_\mathrm{max}$ (and accordingly, the required pump microwave field $b_\mathrm{max}$) becomes very large for large detunings. For example, in our experiment the detuning can far exceed $\Delta = 100\Gamma \approx 200$~MHz, and, accordingly, $b_\mathrm{max} \approx 1.4$~mT, which surpasses typical experimental capabilities. This poses a major challenge for accessing the high-amplitude regime and realizing broadband MFCs.

The nonlinear interaction between the uniform magnon mode and parametrically excited spin waves provides a pathway to overcome this limitation. Driven by a strong off-resonance pump at $\omega_s$, the uniform magnon mode excites pairs of spin waves with wavevectors $\pm \mathbf{k}$ at the same frequency via the second-order Suhl process \cite{1957_Suhl} (Fig.~\ref{fig1}e). As the population of these parametric magnons $N_p  =\sum_{\mathbf{k}}|c_{\mathbf{k}}|^{2}$ grows, the uniform magnon mode experiences a nonlinear frequency shift $\omega_0^N = \omega_0 + \omega_M N_0 + 2 \omega_M N_p$. This shift reduces the effective pump detuning, increasing the uniform magnon mode population $N_0 =|a|^2$, and, consequently, further amplifying the parametric excitation  process. Through this positive feedback loop, the uniform magnon mode amplitude $N_0$ rises rapidly (Fig.~\ref{fig1}f), accompanied by a rapid frequency shift until $N_0 = N_\mathrm{lim}$, where $\omega_0^N$ exceeds $\omega_s$. At this point, the entire magnon band moves above the driving frequency, terminating the parametric process, as indicated by the sharp drop of parametric magnons (red curve) in Fig.~\ref{fig1}f. This threshold behavior---where the uniform magnon population grows explosively---occurs when $N_0$ surpasses the parametric threshold $N_\mathrm{par} = \Gamma / \omega_M$, effectively reducing the high-amplitude transition power from $P_\mathrm{max}$ to a threshold that is close to the parametric excitation threshold $P_\mathrm{par}$. For large detunings ($\Delta > 100\Gamma$), $P_\mathrm{par}$ corresponds to a drive microwave field $b_\mathrm{par} \approx 0.37~\mathrm{mT}$, which is well within reach of our experimental setup.

\begin{figure*}[tb]
\includegraphics[width=0.9\linewidth]{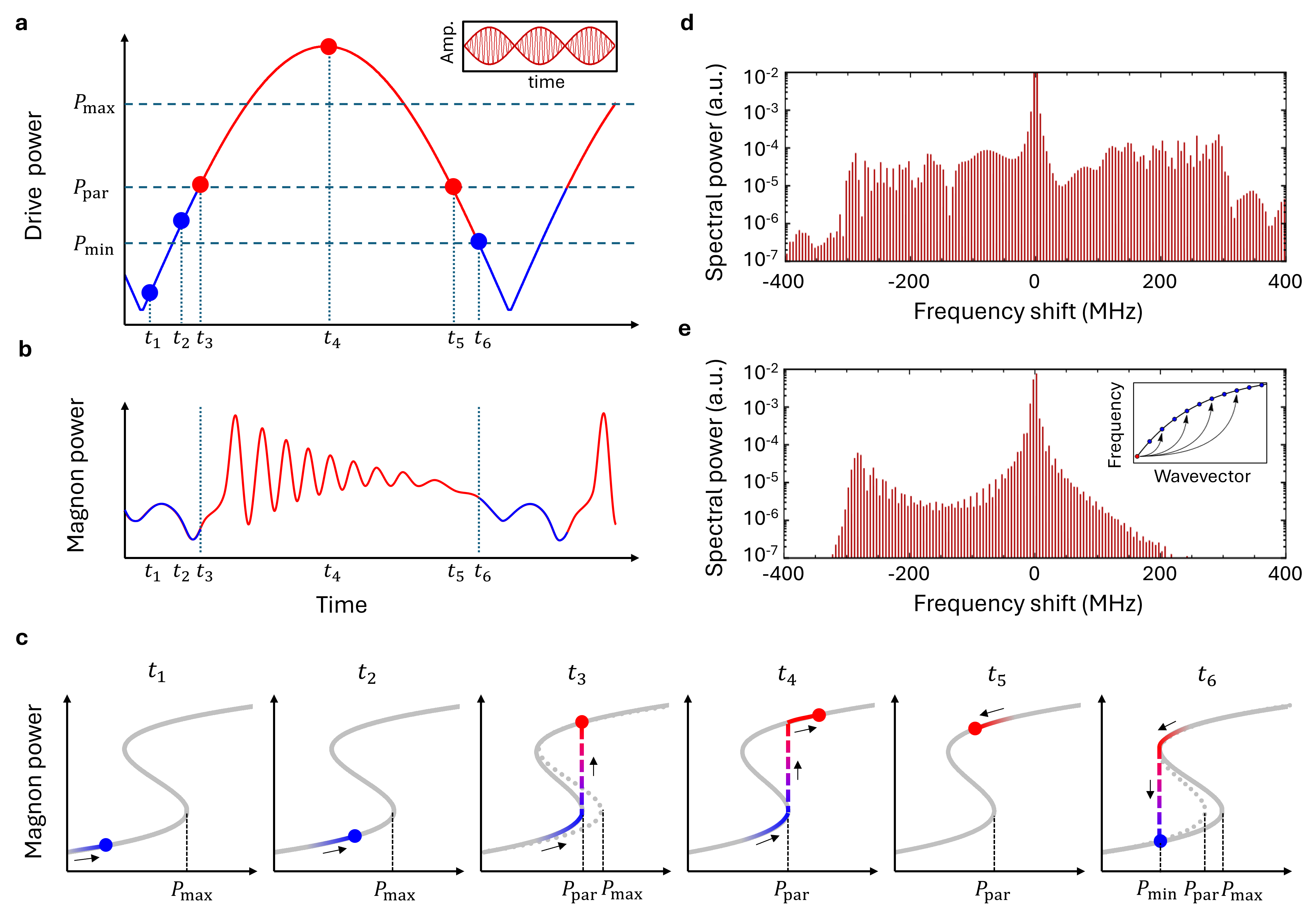}
\caption{\textbf{Dynamic process of magnon frequency comb generation.}
    \textbf{a}, Time trace of the microwave drive signal. $P_\mathrm{min}$, $P_\mathrm{max}$, and $P_\mathrm{par}$ correspond to the turning-point powers shown in Fig.~\ref{fig1}d.
    \textbf{b}, Time trace of the uniform magnon mode power. Amp.: amplitude.
    \textbf{c}, Instantaneous states of the system during one modulation cycle at large detuning. Blue dots and lines: lower branch; red dots and lines: upper branch; arrows indicate sweep direction.
    \textbf{d}, Calculated spectrum of the magnon temporal response under periodic modulation.
    \textbf{e}, Calculated spectrum with the nonlinear dissipation via coupling to higher-order magnon modes accounted for.
    }
\label{fig2}
\end{figure*}

\section{Dynamics of Ultra-broadband Frequency Comb Generation}

Based on these nonlinear processes, the two-tone microwave pump (Fig.~\ref{fig2}a, inset) drives the uniform magnon mode, taking the system through a cycle of bistable transitions (Fig.~\ref{fig1}c). At low drive powers ($t_1$ and $t_2$), the amplitude of uniform magnons $N_0$ follows the lower branch until the positive feedback described above reduces the transition power to $P_\mathrm{par}$ at $t_3$, triggering a jump to the upper branch and a rapid increase in amplitude. The system remains in this high-amplitude state until the periodically varying drive decreases and the magnon power falls to $P_\mathrm{min}$ at $t_6$, returning to the lower branch. With reduced magnon power, the upper transition point resets to $P_\mathrm{max}$. Under periodic modulation, this cycle repeats, producing sustained oscillations in magnon power (Fig.~\ref{fig2}b). The transient decaying oscillations observed after the jump (from $t_3$ to $t_6$) are characteristic of a forced nonlinear oscillator.

This temporal evolution produces an ultra-broadband frequency comb in the calculated spectrum (Fig.~\ref{fig2}d), spanning over 1~GHz, and comprising hundreds of comb lines. The broad spectral width originates from the sharp transition between the lower and upper branches of the bistable response. In our device, this transition is exceptionally abrupt---approaching step-like behavior---due to the large magnon amplitude achieved under the large detuning. Such rapid switching introduces high-frequency components into the temporal dynamics, which translates into a comb extended in the frequency domain. In practical implementations, however, nonlinear coupling between the uniform magnon mode and higher-order spin-wave modes must be considered. These modes, located at higher frequencies (inset of Fig.~\ref{fig2}e), act as additional loss channels that limit the effective bandwidth of the generated comb. Due to this effect, the MFC spectrum becomes asymmetric with respect to the driving frequency, with the majority of comb lines concentrated below $f_s$.

\begin{figure*}[tb]
\includegraphics[width=0.9\linewidth]{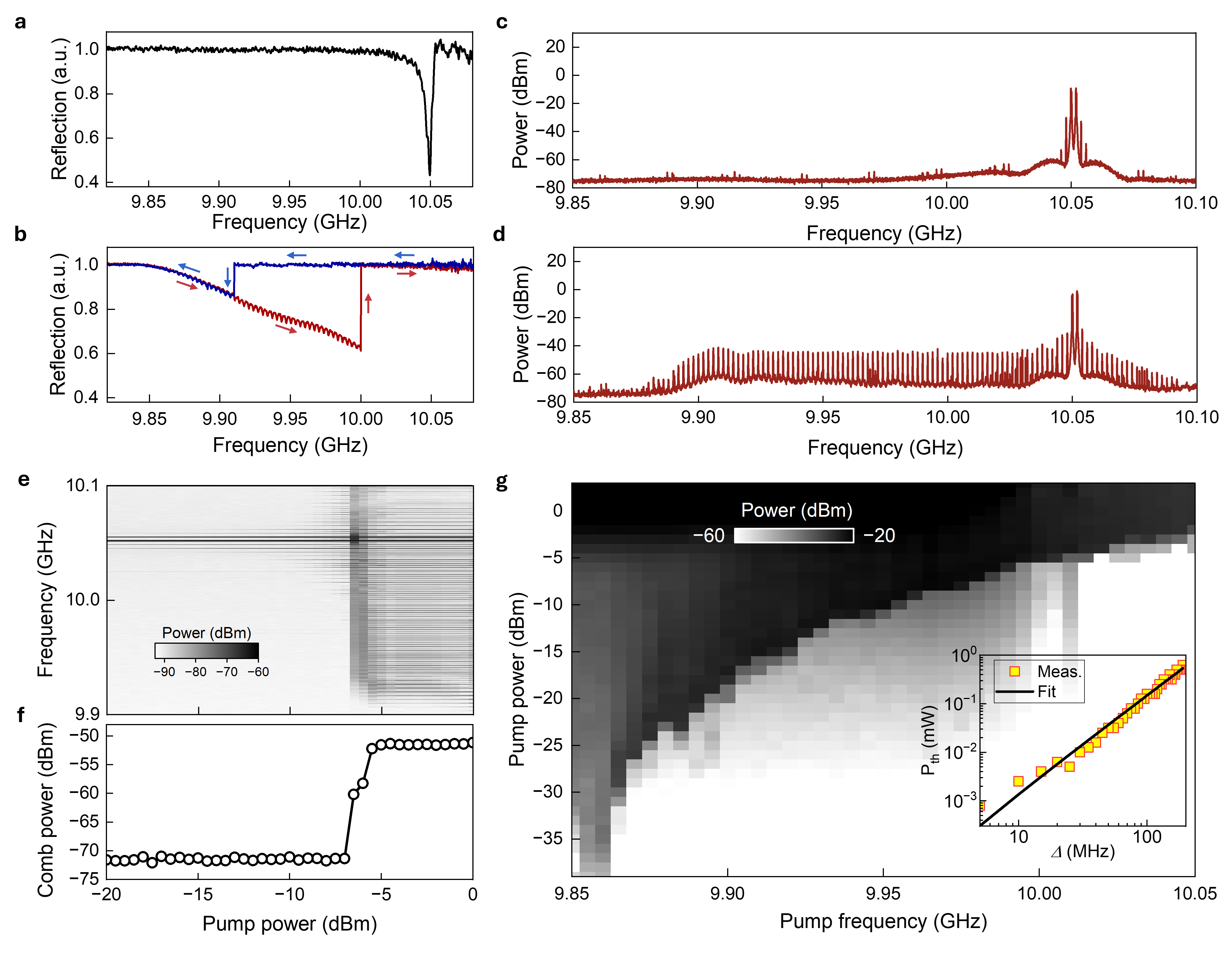}
\caption{\textbf{Experimental spectrum of the MFC.}
    \textbf{a,b}, Device transmission spectra measured at low power ($-30$~dBm) and high power ($-7$~dBm), respectively. The high-power spectrum shows pronounced hysteresis, absent in the low-power regime. Periodic oscillations at high power originate from bulk acoustic phonon modes, a common feature in thin-film YIG magnonic devices \cite{2021_PRAppl_Xu,2025_PRL_Xu}.
    \textbf{c}, MFC spectrum measured with a pump power of $P_s=-30$~dBm at $f_1 = f_0 = 10.05$~GHz.
    \textbf{d}, MFC spectrum measured with a pump power $P_s=-3$~dBm at $f_1 = 10.05$~GHz and $f_0 = 9.86$~GHz.
    \textbf{e}, Output spectra recorded at varying pump powers.
    \textbf{f}, Device output power versus pump power, showing a threshold at $-7$~dBm.
    \textbf{g}, Total MFC power as a function of pump power $P_s$ and frequency $f_1$. Yellow dots indicate extracted threshold points; blue curve shows quadratic fitting.
    }
\label{fig3}
\end{figure*}

\section{Experimental Demonstration of the Ultra-Broadband Frequency Comb}

In our experiments, the MFC device is driven by a microwave pump formed by coherently combining two signals at frequencies $f_1 = f_s = 10.050~\mathrm{GHz}$ and $f_2 = f_s + \delta = 10.052~\mathrm{GHz}$, each with an amplitude of $P_s = -3~\mathrm{dBm}$. The pump frequencies are fixed just below the cutoff of the slow-wave waveguide to maximize slow-wave enhancement, while different pump detuning $\Delta$ is achieved by tuning the uniform magnon mode frequency $f_0$. Fixing the pump frequency rather than the magnon frequency is critical because it ensures that the pump power after slow-wave enhancement (which is strongly frequency-dependent) remains constant across detuning conditions, enabling a fair comparison of comb generation under different detunings. We examine two detuning conditions by biasing the magnon mode with different magnetic fields: (i) small detuning, $f_0 = 10.050~\mathrm{GHz}$ ($\Delta = 0$); and (ii) large detuning, $f_0 = 9.860~\mathrm{GHz}$ ($\Delta = 190~\mathrm{MHz}$). For the small-detuning case, a low sweep power of $-30~\mathrm{dBm}$ reveals the intrinsic magnon resonance as a narrow absorption dip with a linewidth of about $2.0~\mathrm{MHz}$ at $10.05~\mathrm{GHz}$ (Fig.~\ref{fig3}a). In contrast, the large-detuning case uses a higher sweep power of $-7~\mathrm{dBm}$, producing a broad resonance spanning $170~\mathrm{MHz}$—nearly two orders of magnitude wider than the magnon linewidth—and clear hysteresis between upward and downward sweeps (Fig.~\ref{fig3}b). The span of this nonlinear resonance increases further with higher power and can eventually cover the pump frequency in our experiment.

In both cases, MFCs with the comb line spacing equal to $\delta = 2$~MHz are produced, but their spectral characteristics differ significantly. In the small-detuning case, the resonance shifts away due to the strong nonlinear effect under a strong pump, limiting the magnon amplitude and the conresponding comb bandwidth. This prediction matches the measured MFC in Fig.~\ref{fig3}c, which spans only 10~MHz with six comb lines. In the large-detuning case, the comb spans 9.87 -- 10.09~GHz (Fig.~\ref{fig3}d), corresponding to 220~MHz and about 100 comb lines, which is an order-of-magnitude increase. The comb shows a relatively flat envelope and pronounced asymmetry around the driving frequency, with more than 70 comb lines below and fewer than 20 above. The typical power per comb line is about $-45$~dBm ($\approx 30$~nW). All reported power values refer to on-device levels after accounting for input and output losses.

As predicted by our theoretical analysis, MFC generation exhibits a clear threshold behavior. Figure~\ref{fig3}e shows the measured output spectra with the input pump power ($P_s$) sweeping from $-20$ to $0$~dBm, and Fig.~\ref{fig3}f plots the extracted total comb power versus input power. A distinct threshold appears at $P_s = -7$~dBm, where the output power abruptly increases by about 20~dB. Beyond this point, the output rapidly evolves into a broadband MFC with a bandwidth exceeding 200~MHz. Furthermore, the output power saturates above the threshold, consistent with the theoretical prediction in Fig.~\ref{fig2}c, which indicates that the comb originates from periodic sharp transitions between the lower and upper branches of the bistable nonlinear magnonic resonance curve, while the power variation within the upper branch after the transition play only a minor role. Our analysis also predicts that the threshold power depends on the driving frequency detuning $\Delta$, which is confirmed by the measured comb power (color coded in Fig.~\ref{fig3}g) as a function of both the pump power and pump frequency. When $f_s < f_0$, no MFC generation occurs. When $f_s$ exceeds $f_0$, a clear input power threshold emerges, above which the comb generation begins. The extracted threshold power $P_\mathrm{th}$ is plotted as a function of $\Delta$ in the inset of Fig.~\ref{fig3}g (yellow squares), which follows a parabolic relation according to our numerical fitting (solid blue line), in excellent agreement with theoretical predictions.

\begin{figure*}[tb]
\includegraphics[width=0.9\linewidth]{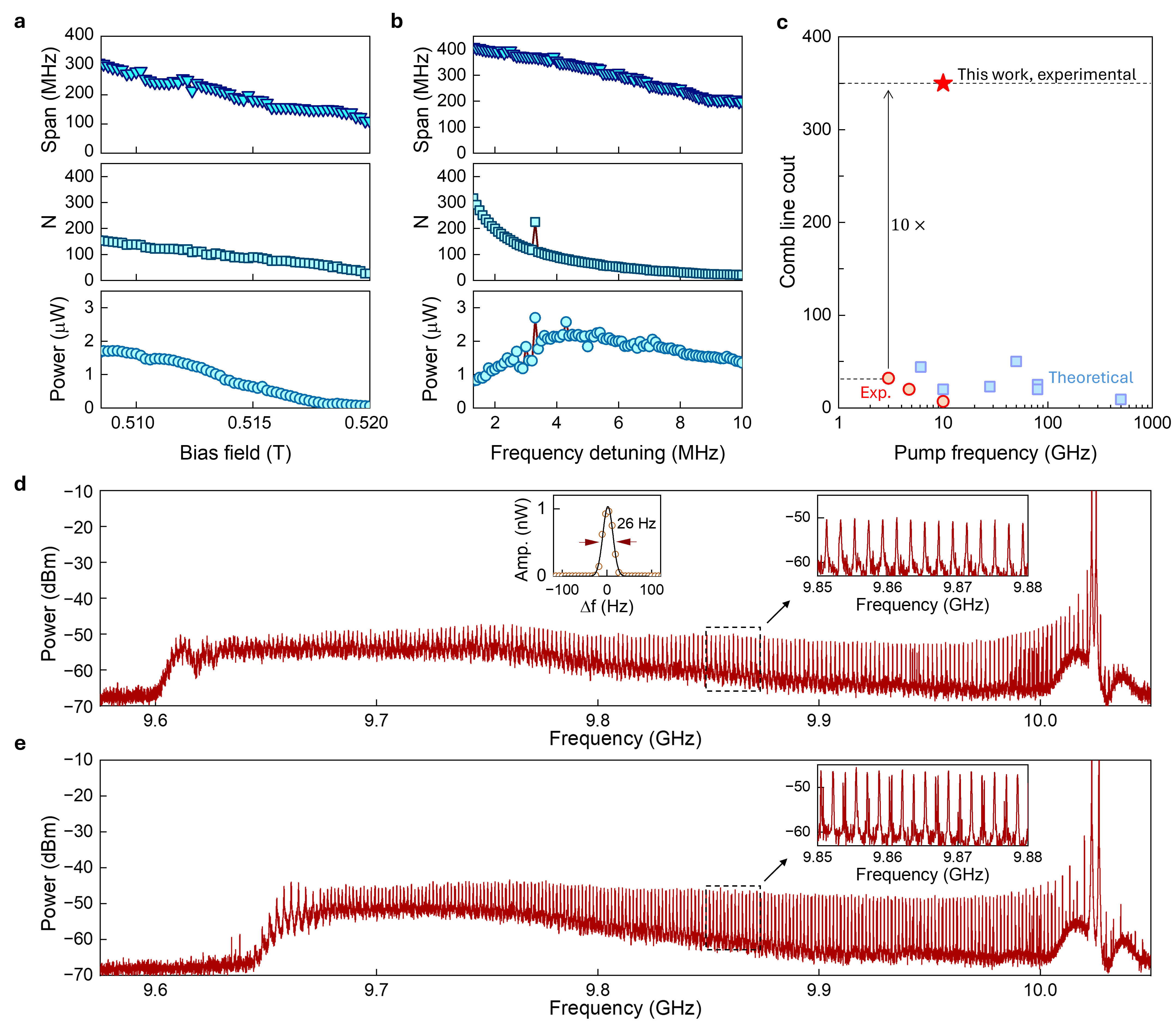}
\caption{\textbf{Tuning of the MFC bandwidth.}
    \textbf{a,b}, MFC characteristics--span, line count, and total on-chip comb power--as functions of the bias magnetic field and pump frequency difference $\delta$, respectively.
    \textbf{c}, Comparison of comb line count achieved in this work (red star) with state-of-the-art results, including experimental demonstrations (red circles) and theoretical proposals (blue squares), extracted from Refs.\cite{2024_NP_Wang_EPMFC,2023_PRL_Xu_MagnomechanicalResonator,2022_APL_Hula_SpinWaveFC,2025_APLQuantum_Kani_SqueezedComb,2023_FundResearch_Xiong,2023_PRA_Liu_TwoToneDrive,2025_PRA_Wang_MechanicalMFC,2022_PRL_Wang_Twisted,2024_NanoLett_Li_AsymmetricMFC,2021_PRL_Wang_nonlinearMagnonSkyrmion,2021_AdvEngMatt_Sun_StrainModulation,2024_PRB_Liu_SyntheticFerrimagnets}. Exp.: Experimental results.
    \textbf{d}, Measured MFC spectrum with a comb line spacing of 2~MHz and a total of 200 lines, obtained with uniform magnon mode frequency $f_0 = 9.35$~GHz, pump frequency $f_1 = 10.0232$~GHz and $f_2=10.0252$~ GHz, and pump power $P_s = -3$~dBm. Right inset: magnified view of the comb spectrum. Left inset: magnified view of a single comb line with Gaussian fitting, revealing a linewidth of 26~Hz.
    \textbf{e}, MFC spectrum exhibiting period doubling, obtained with uniform magnon mode frequency $f_0 = 9.35$~GHz, pump frequency $f_1 = 10.0232$~GHz and $f_2=10.0265$ GHz, and pump power $P_s = -3$~dBm. The comb line spacing is $\delta/2 = 1.65$~MHz, with a total line count of 225. Inset: magnified view of the comb spectrum.
}
\label{fig4}
\end{figure*}

\section{Magnonic Frequency Comb Bandwidth Expansion }

The versatility of magnons enables flexible control of the MFC properties through multiple parameters for a given device. Owing to the magnetic tunability of the magnon frequency and the design flexibility of the slow-wave waveguide, MFCs can be generated across different frequency bands. In our experiments, ultra-broadband MFCs have been reliably achieved within the 3--10~GHz range. Beyond the operating frequency, key MFC characteristics---including the comb span, the number of comb lines, and the total comb power (excluding pump tones)---can be precisely controlled by adjusting the bias magnetic field. Figure~\ref{fig4}a shows MFC parameters obtained at different bias magnetic fields under fixed pumping conditions: $f_1 = 10.005$~GHz, $\delta = 2$~MHz, and $P_s = -3$~dBm. As the bias field decreases, the corresponding reduction in the magnon frequency $f_0$ increases the pump detuning $\Delta$, which, in turn, broadens the MFC span to as much as 300~MHz with approximately 150 comb lines. Under these conditions, the total on-chip comb power reaches 1.8~$\mu$W, corresponding to an average power of about 12~nW per comb line.

The MFC properties can also be effectively controlled by the pump spacing $\delta$, as shown in Fig.~\ref{fig4}b. For instance, the comb span increases monotonically as $\delta$ decreases. When $\delta = 1.3$~MHz, the comb span extends to 450 MHz with a total of 350 comb lines. However, the dependence of the comb power on $\delta$ is non-monotonic. As $\delta$ decreases from 10~MHz to 4.3~MHz, the comb power increases. Beyond this point, although the comb line count continues to grow, the power per line diminishes, leading to a reduction in total comb power. The maximum total on-chip comb power achieved in our experiments is 2.58~$\mu$W at $\delta = 4.3$~MHz, corresponding to an average of 31~nW per comb line.

The 350 comb line count obtained in our experiment represents the highest achieved to date in MFCs. Figure~\ref{fig4}c provides a summary of recent progress in this field, serving as a benchmark for comparison. The comb-line count reported in this work exceeds all previous experimental demonstrations by at least one order of magnitude and even far surpasses existing theoretical predictions and simulations. This establishes our ultra-broadband MFC platform as the most advanced among reported microwave MFC systems, regardless of operating frequency, material platform, or physical mechanism. A parallel study \cite{2025_PRL_Xu} was reported independently during the preparation of this manuscript, which demonstrated an MFC with 130 comb lines by exploiting a series of equidistant, low-loss bulk acoustic resonances in the substrate of the YIG-film resonator. This approach relies on the substrate and imposes a fixed comb spacing determined by the FSR of the mechanical resonances. Our method is fundamentally different: it does not depend on the mechanical resonance in the substrate and can, in principle, be implemented on arbitrary substrates, offering great versatility for on-chip integration. Furthermore, our approach enables continuous tuning of comb line spacing, a unique advantage of MFCs compared with other frequency comb platforms.

Figures~\ref{fig4}d show the spectrum of the widest MFC obtained in our experiments along with a close-up view of its comb lines, clearly revealing individual lines spaced by 1.3~MHz. Gaussian fitting yields a single comb-line linewidth of $26$~Hz (left inset of Fig.~\ref{fig4}d), limited by the resolution bandwidth of the spectrum analyzer, indicating excellent coherence with an actual linewidth below this value. The comb lines are superimposed on a broadband continuum background, which is likely attributed to spontaneous four-magnon scattering from the pump tones \cite{2022_APL_Hula_SpinWaveFC}. This magnon continuum is more pronounced at the lower end of the MFC band, where it lies closer to the intrinsic frequency of the uniform magnon mode. Consequently, the extinction ratio of the comb lines is higher at the upper end of the span, and gradually decreases toward the lower frequency end.

Another practical approach to further increase the comb line count is to exploit the period-doubling effect, a well-established feature in frequency comb systems. For a given MFC, this phenomenon typically occurs prior to the onset of chaotic dynamics, and it can approximately double the number of comb lines. In our system, this effect is observed at $\delta = 3.3$~MHz. As shown in Fig.~\ref{fig4}e, it results in a comb line count of 225 with a halved line spacing of $1.65$~MHz.

\section{Discussion and outlook}

We present a novel approach for realizing ultra-broadband MFCs by engineering the nonlinear response of a nonlinear magnonic YIG-film microresonator. Leveraging the interplay between nonlinear frequency shift and parametric pumping, our platform achieves low-threshold comb generation with a bandwidth of up to 450~MHz and a comb line count of 350. To the best of our knowledge, this represents the highest experimentally reported magnonic comb-line count, exceeding previous demonstrations by more than an order of magnitude and surpassing existing theoretical proposals (Fig.~\ref{fig4}c). The MFC properties---including bandwidth, line density, and power distribution---can be tuned via external parameters such as magnetic bias and pump detuning, revealing rich nonlinear dynamics and enabling precise control of comb structures through bistability engineering.

Beyond performance metrics, this work establishes a scalable, hardware-efficient platform for integrating frequency comb functionality into magnonic circuits. The spectral richness of MFCs enables access to high-dimensional mode spaces, which can be harnessed for advanced computing and sensing applications. In particular, the inherent parallelism of the comb structure supports operations such as vector–matrix multiplication \cite{2021_Nature_Xu_PhotonicTensor,2022_Light_Zhou_PhotonicMatrix,2024_Optica_Latifpour,2025_NatElectron_Ji_SyntheticDomain}, a key primitive in the neuromorphic and analog computing. These capabilities are achieved without increasing the physical complexity or footprint of the device, by exploiting the frequency domain as an additional degree of freedom. By uniting compactness, tunability, and computational potential, our MFC platform offers a compelling alternative to earlier established photonic, acoustic and magneto-acoustic systems. Its integration into magnonic architectures opens new directions for microwave signal processing, high-dimensional information processing, and precision sensing, laying the groundwork for next-generation functional devices driven by nonlinear magnetic phenomena.

\section{Methods}
\textbf{Device description}: The magnonic micro-resonator is a 200-nm yttrium iron garnet (YIG) thin film patterned into a $50~\mu m \times 50~\mu m$ square on a 500-$\mu$m-thick gadolinium gallium garnet (GGG) substrate. The YIG chip is biased by an out-of-plane magnetic field, supporting magnons in the forward magnetostatic volume mode. This YIG structure is flip-bonded onto the slow-wave waveguide, which consists of a periodically corrugated microstrip fabricated on a printed circuit board (PCB). Microwave signals propagating along the corrugated microstrip excite spoof surface plasmon polaritons, which exhibit significantly reduced group velocity and enhanced mode confinement. This leads to a substantial enhancement in the pumping efficiency of magnon modes compared to conventional transducers such as microstrips or coplanar waveguides (CPWs), thereby boosting the nonlinear dynamics in the YIG resonator. Notably, due to the propagating nature of spoof surface plasmon polaritons, this enhancement is broadband, which is crucial for the broadband generation of MFCs.

\bibliography{comb_bib}

\end{document}